\documentclass[preprint,showkeys,preprintnumbers,amsmath,amssymb]{revtex4}

\usepackage{graphicx}
\usepackage{dcolumn}
\usepackage{bm}

\begin{document}

\preprint{APS/123-QED}

\title{Metallic coatings of MEMS at low temperatures:\\ stress, elasticity and non-linear dissipation}

\author{E. Collin} 
\email{eddy.collin@grenoble.cnrs.fr}
\author{J. Kofler}
\author{S. Lakhloufi}
\author{S. Pairis}
\author{Yu. M. Bunkov}
\author{H. Godfrin}

\affiliation{%
Institut N\'eel
\\
CNRS et Universit\'e Joseph Fourier, \\
BP 166, 38042 Grenoble Cedex 9, France \\
}%

\date{\today}

\begin{abstract}
We present mechanical measurements performed at low temperatures on cantilever-based micro-electro-mechanical structures (MEMS) coated with a metallic layer. Two very different coatings are presented in order to illustrate the capabilities of the present approach, namely (soft) aluminum and (hard) niobium oxide. The temperature is used as a {\it control parameter} to access materials properties. We benefit from low temperature techniques to extract a phase-resolved measurement of the first mechanical resonance mode in cryogenic vacuum. By repeating the experiment on the same samples, after multiple metallic depositions, we can determine accurately the contribution of the coating layers to the mechanical properties in terms of surface stress, additional mass, additional elasticity and damping. Analytic theoretical expressions are derived and used to fit the data. Taking advantage of the extremely broad dynamic range provided by the technique, we can measure the anelasticity of the thin metallic film. The key parameters describing the metals' dynamics are analyzed in an original way in order to provide new experimental grounds for future theoretical modelings of the underlying mechanisms. 
\end{abstract}

\maketitle

\section{INTRODUCTION}

Micromachined 
 mechanical devices \cite{cleland} have attracted the interest of physicists and engineers for decades, and their field of research and applications is continuously expanding. 
Devices based on micro-electro-mechanical systems (MEMS) are commercially available for a wide range of applications (see e.g. \cite{mems}), with in particular topical developments in microfluidics \cite{fluids} and chemical sensing \cite{sensing1,sensing2}.

Low temperature physics and techniques is a field were microfabricated mechanical devices have a role to play in the future.
For instance, reliable and if possible scalable and versatile 'micro-viscometers' are of great importance to low temperature physicists, who used so far a wide variety of objects: metallic microspheres, hand-made vibrating wires and vibrating mesh-grids for the study of quantum turbulence in both $^4$He and $^3$He \cite{sphere4,turb4,turb3}, vibrating wires for the definition of an ultra-low temperature scale and its practical measure in $^3$He \cite{lancaster3Hethermo,us3Hethermo}, and vibrating wires again for ultra-sensitive bolometry at 100$~\mu$K applied to particle detection \cite{particleLancaster,bolo}. 

In the attempt to develop better probes, commercially available quartz tuning forks are nowadays used in many low temperature laboratories \cite{fork1,fork2}. The Grenoble group \cite{trique00} on the other hand started to use microfabricated silicon MEMS devices to replace advantageously the 'classical' vibrating wire technique.

Indeed silicon is perfectly fit to this use, due to its small low temperature mechanical dissipation \cite{parpiadissipeMEMS}, high yield strength, small surface roughness and a large panel of fabrication techniques enabling complex designs. 

These micro-devices are usually used in a {\it resonant} way in order to improve their sensitivity through the quality factor $Q$ of the mechanical mode. This quality factor is a signature of the dissipation mechanisms occurring in the devices. Understanding these mechanisms is thus of great importance to many communities using micro and nowadays nano mechanical structures, from gravitational wave detection \cite{gravit} to the appealing possibility of controlling the {\it quantumness} of a nano-beam \cite{quantbeam}. 
Furthermore, identifying the dominant dissipation mechanisms is a fundamental issue, tackling for instance thermoelastic damping \cite{thermodamp} and the tunneling of two-level systems, characteristic of glasses \cite{glassTLS}.

The mechanical motion of these low temperature devices is induced and/or detected by electric means, and conducting layers are incorporated in their design. 
Thin metallic coatings impact the mechanical properties in a way that deserves to be fully understood. It is the subject of the present article.

Of course, our low temperature characterization of mechanical properties of thin metallic films transcends the somewhat peculiar low temperature field. 
A film of any substance deposited on a cantilever will impact its static and dynamic response. This fact is of great importance for micro-electronics since most components are made of layered structures, which reliability and lifetime is determined, at least in part, by their mechanical properties \cite{hu-components}. Moreover, cantilever-based devices can be used to {\it characterize} the properties of a given coating. In a static mode, this is the well known problem first solved by Stoney \cite{purestoney} (and widely adapted since) applied to the determination of surface stresses \cite{stoneyfact,kleinstoney,sadercantilever2001,ZhangZhao1,ZhangZhao2}. In a {\it dynamic mode}, it was first pointed out that a mechanical resonance frequency is extremely sensitive to additional mass deposition (originally with Quartz Crystal Microbalances, QCM) \cite{massdetect}, and also to the elasticity of an additional layer \cite{Petersen}. Recently the effect of the surface stress on the mechanical resonance modes has been 
considered \cite{oshea,mcfarlandsurfacestress}, pointing out that a mechanical mode frequency is also strongly stress-dependent.

In our experiments, we {\it benefit from low temperature techniques}. We maintain cryogenic vacuum around the samples (i.e. below $10^{-6}~$mbar), and use a standard low-noise magnetomotive detection scheme. Moreover, the mechanical properties improve at low temperatures ensuring high $Q$ values (up to $0.3~10^6$ at 4$~$K). The temperature turns out to be a {\it control parameter} together with the driving force enabling a comprehensive characterization of the devices. The drive and detection schemes enable very large displacements, the high yield strength of silicon preventing breaking. By performing multiple depositions of {\it the same metal on the same sample}, we can extract the true contribution of the coating to the mechanical resonance properties. 

The aim of the paper is to provide original results on both the theoretical modeling and the experimental data. 
Generic analytic expressions are derived and presented as tools for detailed characterization of the dynamics of metal-coated cantilevered MEMS.
The experimental study is performed from the linear to a highly non-linear regime, with the help of the theoretical input of \cite{JLTP_VIW}. We finally present the key parameters describing the impact of the metal on the dynamics of the silicon MEMS. These new experimental low temperature grounds are calling for further theoretical investigations addressing the microscopic mechanisms at work in the materials. 


\section{EXPERIMENT}

\begin{figure}[t!]
\includegraphics[height=8cm]{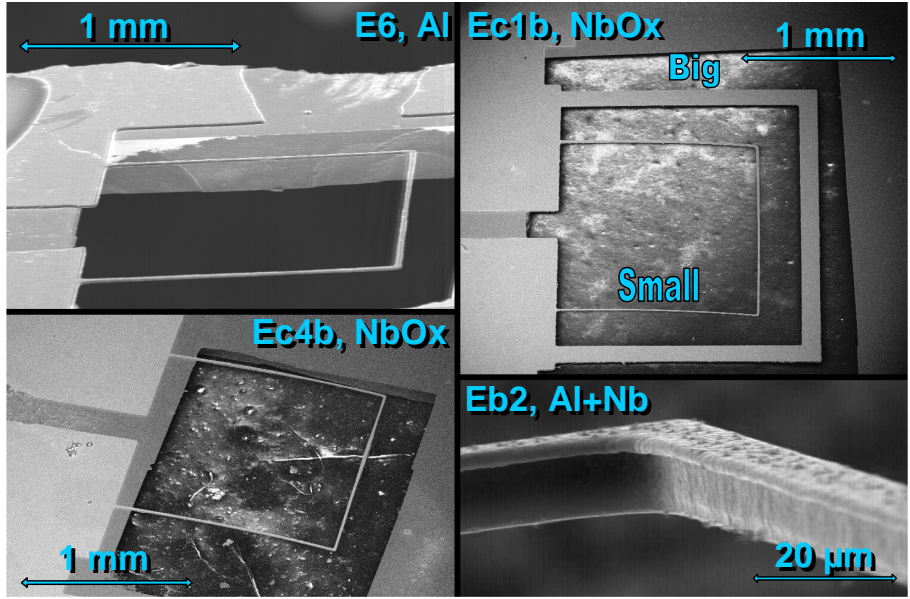}
\caption{\label{pictures} (Color online)  
Goal-post structures realized for our study. Top-left: sample E6, aluminum coated. Bottom-left: sample Ec4b, coated with oxidized niobium. Top-right: a double goal-post structure, realized with an oxidized niobium coating (sample Ec1b, 'big' and 'small' resonators); these two structures are electrically in parallel. Bottom-right: a close-up on one sample (Eb2 with a composite coating, an aluminum and niobium sandwich where the aluminum was used as etching mask).
On these FE-SEM (Field Emission Scanning Electron Microscope) pictures, all structures are single-side coated.
}
\end{figure}

We report on the impact of the surface stress, elasticity and mass of the (normal-state) coating metal on the resonance of cantilever-based silicon MEMS for temperatures between $1.5~$K and $35~$K. 
The signatures of the friction mechanisms present in the metal will be discussed as a function of the devices characteristics.
Anelastic behavior of metals under very high (dynamic) strains is reported.

The experimental setup and fabrication technique have been described in \cite{JLTP_VIW}, and we shall refer extensively to this work in the following. 

\subsection{Samples}

The samples consist in goal-post shaped silicon structures, namely two cantilevers (hereafter called 'feet') clamped together by a beam (called 'paddle'), see Fig. \ref{pictures} and \ref{schematic}. On top of the silicon, a non-superconducting metallic layer is used to drive and detect the motion of the device. 
Two metal coatings are discussed in the present article in order to illustrate the capabilities of our measurement scheme: E6 is (soft) aluminum coated (it is the same sample as in \cite{JLTP_VIW}), and the two others, Ec1b and Ec4b are coated with a (hard) oxidized niobium layer. The aluminum was Joule evaporated in good vacuum at room temperature, while the niobium was magnetron deposited in a small residual air pressure with a water-cooled sample holder. No annealing procedure was performed. For the aluminum-coated samples, the metal was used as mask for the last Reactive Ion Etching (RIE) step (see Fig. \ref{pictures}). The niobium samples used an S1818 masking overlayer, which was removed in an 0$_2$ plasma.
The aluminum thin films are reasonably clean, with a superconducting transition temperature around $1.4~$K \cite{alsuprafilm}, while the niobium films contain oxygen and display a suppressed $T_c$ around $1.6~$K. From the literature \cite{nboxsupra}, about 8.5$~$\% at. O (on interstitial sites) would be necessary for such a decrease. An X-ray analysis of our samples confirmed the presence of approximately 10$~$\% at. oxygen, with only traces of other contaminants \cite{SEBP}.
\begin{figure}[t!]
\includegraphics[height=5cm]{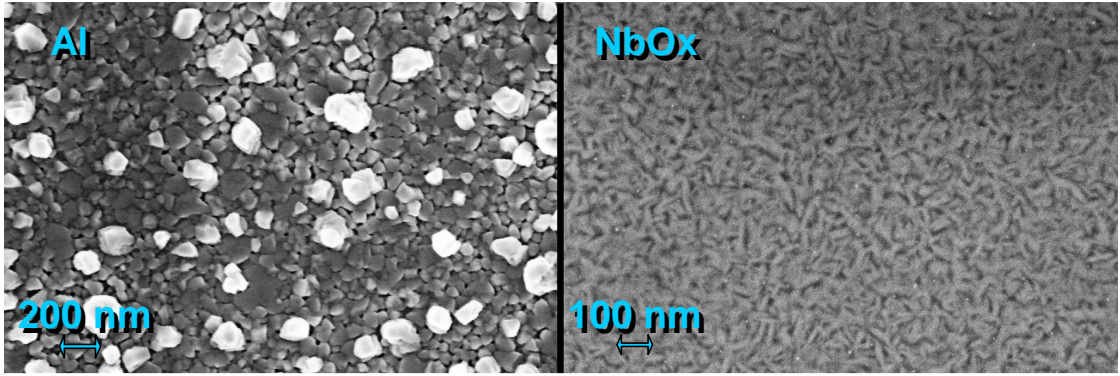}
\caption{\label{surface} (Color online) Left: FE-SEM (Field Emission Scanning Electron Microscope) picture of an aluminum Joule-evaporated layer. Right: FE-SEM picture of a niobium magnetron-deposited layer (in a small residual air pressure). Both are about 200$~$nm thick, but display very different structures.}
\end{figure}
A Scanning Electron Microscope (SEM) picture of the surface coatings is given in Fig. \ref{surface} showing two very different structures: a broad distribution of grain size for our aluminum with a few very large ones (up to 200$~$nm), and a very homogeneous distribution with small grains for our oxidized niobium (slightly anisotropic, but about 50$~$nm on average). The sample Ec1b consists in two goal-post structures electrically in parallel (referred to thereafter as 'big' and 'small'). 

In order to resolve the contribution of the (normal-state) metallic layer to the mechanical resonance, we performed on E6 and Ec4b multiple depositions of the same metal. Moreover, in order to separate the effect of {\it the axial load} from the {\it bending moment} generated both by the surface stress due to the metal, we did these depositions {\it on the two sides of each sample} (hereafter called front-side and back-side). The samples were kept clean, but no special preparation was performed on the surfaces prior to deposition.

\subsection{Setup}

\begin{figure}
\includegraphics[height= 6cm]{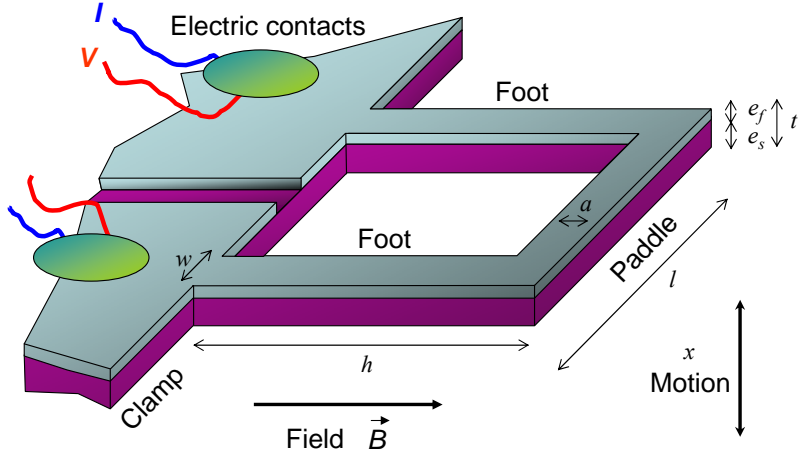}
\caption{\label{schematic} (Color online)  
Schematic drawing of the goal-post structures. The Ec1b sample includes two such structures imbricated one in the other, and connected electrically in parallel. For the drawing, only one side coating is represented (front-side layer, bright color), on top of the silicon (darker color). The drive current $I$ and detection voltage $V$ are represented together with the magnetic field $B$ and the displacement $x$ of the paddle. 
}
\end{figure}

A $^4$He pumped cryostat is used to reach temperatures from 35$~$K down to 1.5$~$K. The samples were placed in a vacuum chamber (P$<10^{-6}~$mbar), on a temperature-regulated copper plate. The drive and detection scheme is based on the vibrating wire technique used in low temperature physics for viscometry  (see e.g. \cite{vibwireold}). A current $I=I_0 \cos(\omega t)$ is fed into the metallic layer, which stands in the center of a coil producing a static field $B$ parallel to the sample surface. A magnetomotive Laplace force $F=I_0 l B \cos(\omega t)$ acts on the paddle of length $l$, driving it out of the sample plane (Fig. \ref{schematic}).
Through Faraday's law the motion induces a voltage $V=l B v \cos(\omega t+\phi)$, with $v$ the speed amplitude of the top part of the structure and $\phi$ its phase. The signals, detected with a lock-in amplifier, are a voltage component $X=l B v \cos(\phi)$ in-phase with the driving force $F$ (or the current $I$), and an out-of-phase one $Y=l B v \sin(\phi)$. As the angular frequency $\omega$ is swept around the first mechanical mode frequency $\omega_0$, the structure is brought to resonance and a peak is detected, Fig. \ref{detect}. The displacement amplitude of the top part of the structure is simply $x=v/\omega_0$, even at very strong excitations.
The drive currents $I_0$ used were always smaller than 100$~\mu$A ({\it rms}) and the fields $B$ smaller than 350$~$mT. No anomalous field (or current) dependence could be detected in this range of parameters. Displacements were typically in the range $0.5~\mu$m - $0.2~$mm ({\it rms} values).

\subsection{Raw data}

\begin{figure*}
\includegraphics[height=6cm]{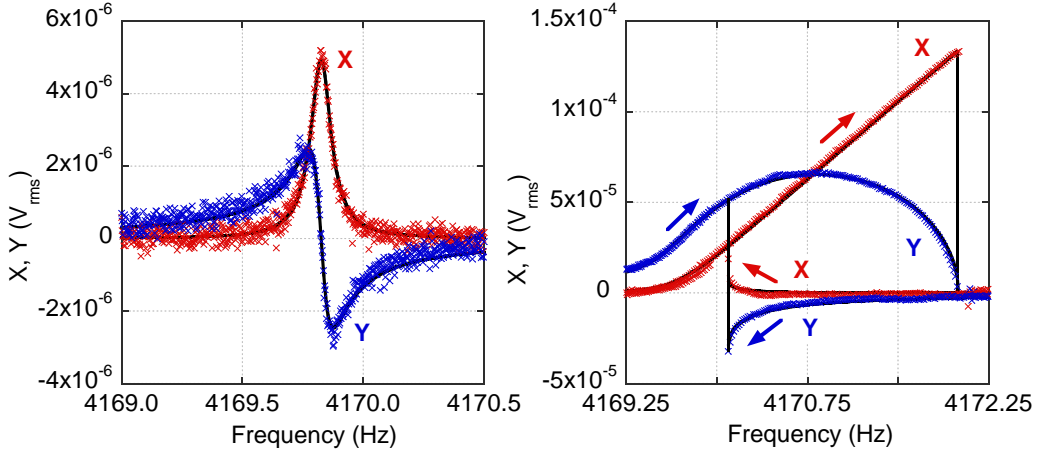}
\caption{\label{detect} (Color online) Signal recorded at the first mechanical mode resonance, for sample Ec4b in vacuum at 4$~$K. $X$ is the voltage in-phase with the driving force while $Y$ is the out-of-phase component. Left: small drives (force 24$~$pN$_{rms}$, displacement 3$~\mu$m$_{rms}$) in the linear regime, the black line is a Lorentzian fit. Right: deeply in the non-linear regime (force 0.6$~$nN$_{rms}$, displacement 78$~\mu$m$_{rms}$). The resonance line shows hysteresis while sweeping the frequency up, or down (arrows). The black lines are fits following the theory of \cite{JLTP_VIW}. Both curves are taken in a 50$~$mT field.}
\end{figure*}

We showed in \cite{JLTP_VIW} that in our experiments even for the largest drives, the structure can be thought of as two cantilevers (the feet) perfectly clamped together, moving together in the common motion of the first mechanical mode.   
At small excitations, the line is {\it perfectly Lorentzian} (Fig. \ref{detect}, left). From the fit one extracts an amplitude $V_{max}$ (the height of the $X$ component), a linewidth $\Delta f$ (the full width at half height of $X$) and a resonance frequency $f_0=\omega_0/2 \pi$ (the position of the maximum of $X$, and the zero on $Y$). 
One can easily show that $V_{max}$ and the force $F_0=I_0 l B$ are linked together through simple relations:
\begin{eqnarray}
 V_{max} & = & l B \omega_0 \, x_{max}, \label{Vmaxdef}\\
 x_{max} & = & Q \, F_0 /k \,  ,  \label{displ}
\end{eqnarray}
 with 
\begin{equation}
 Q=f_0/\Delta f \label{Qdef}
\end{equation}
the quality factor of the resonance, and $x_{max}$ the maximal deflection amplitude at resonance. In (\ref{displ}) $k$ is the spring constant of the mode, and 
\begin{equation}
\omega_0=\sqrt{k/m} \label{omegadef}
\end{equation}
gives access to the mass $m$ associated to the mode.
However at large drives, the resonance line becomes non-linear (Fig. \ref{detect}, right). At a critical value of the displacement $x_c$, the curve is bi-valued and the measurement shows hysteresis \cite{landaumeca}. We demonstrated that for our structures, the global shape of the resonance line is due to a geometrical non-linearity which pulls up the resonance frequency as $f_r= f_0 +\beta \, x_{max}^2$ for an upwards frequency sweep (with $\beta>0$ a temperature-independent coefficient characteristic of the structure) \cite{JLTP_VIW}. From the fits one can still extract the 'linear' resonance frequency $f_0$ (with $f_r$ the position of the maximum of $X$ and zero of $Y$ in an up-sweep), the height $V_{max}$ (the height of the $X$ component when the frequency is swept upwards) and the 'linear' linewidth $\Delta f$ which is {\it not} the width of the resonance curve. One can show that $V_{max}$, $f_0$ and $\Delta f$ still follow Eq. (\ref{Vmaxdef})-(\ref{omegadef}).

\section{THEORETICAL TOOLS}
\label{theor}

From the measurement we obtain $V_{max}$, $f_0$ and $\Delta f$ as a function of the experimental conditions, namely temperature and driving force. A full series of measures is performed for each sample, for each metallic layer deposition. 

From Eq. (\ref{Vmaxdef})-(\ref{omegadef}) we then extract $m$, the vibrating mass of the mode, and $k$ its spring constant.
The {\it friction mechanisms} present inside the structure are responsible for the linewidth $\Delta f$ (dissipative contribution), but they also impact the frequency through a shift $\delta f$ (reactive contribution). Both are linked through Kramers-Kronig relations valid in the linear regime, since they are related to the real and imaginary parts of the acoustic susceptibility of the cantilevers \cite{castroNeto}.
The friction can be described by a viscous force $F_{friction}= -2 \Lambda_2 \, \dot{v} -2 \Lambda_1 \, v$, leading to:
\begin{eqnarray}
\delta \omega & = & - \frac{\omega_0  \Lambda_2}{m}, \\
\Delta \omega & = & 2\, \frac{ \Lambda_1}{m},
\end{eqnarray}
expressed in rad/s. Experimentally, the shift (and thus $\Lambda_2$) is included in the measured spring constant $k$.

$m$, $k$ and $\Lambda_1$ are the {\it only parameters} (together with the geometrical non-linear coefficient $\beta$) required to fully describe the resonance. However, at very large excitations the materials can display {\it anelastic behaviors}. This means in general that both the measured spring constant $k(x)$ (through $\Lambda_2(x)$) and the dissipation constant $\Lambda_1(x)$ will be functions of the displacement $x$. This point has to be taken into account to fit the resonance lines; also, the properties 'at resonance' become functions of the maximal deflection amplitude, and our discussion will be dealing with  $k(x_{max})$ and $\Lambda_1(x_{max})$. Note that in our modeling the vibrating mass of the mode $m$ is independent of the stress state of the structure. 

We now turn to the theoretical modeling of the beams dynamics which enables to interpret $m$ and $k$. The linewith parameter $\Lambda_1(x)$ will be correlated to these results and discussed in the last parts of the paper.

\subsection{Harmonic parameters}

For long and thin cantilever beams, the Euler-Bernoulli 1D equation is perfectly adequate \cite{eulerB}. From \cite{JLTP_VIW,timovibr} we write for the first mode of the bare silicon structure:
\begin{eqnarray}
k & = &  2 \left(\frac{E_s \, I}{h^3} \right)  \, \lambda^4 \, \left[  \gamma_{0} +  \frac{\eta}{2} \right], \label{kspring}\\
m & = &  2 \, (\rho_s \, t w h) \left[ \gamma_{0} + \frac{\eta}{2} \right],  
\end{eqnarray}
with $E_s$ the Young modulus of silicon along the beam and $I=w t^3/12$ the flexural moment of inertia. $\rho_s$ is the density of silicon, and $h$ the length of the feet, $w$ the width of the feet and $t$ their thickness (Fig. \ref{schematic}). These expressions are obtained by solving the harmonic problem, and making use of the method of the virtual work. $\lambda$ and $\gamma_0$ are mode-dependent constants (given here to about 0.5 \% for the first mode):
\begin{eqnarray}
\lambda(u,\eta) & = & \lambda_0(\eta) \left( 1 - \frac{u}{2} \, \phi_0(\eta) \right), \label{lambda}\\
\lambda_0(\eta) & = & \left[ \frac{(\lambda_{00})^4/4 + 3 \, \eta/2 }{1 + \eta/2} \, \frac{1}{1/4 + \eta/2}\right]^{1/4} ,\\
\gamma_0 (\eta) & = & \frac{\frac{1}{4} +  \frac{33}{140} \,\, 4.418 \, \eta/2}{1 + 4.418 \, \eta/2} , \\
\phi_0(\eta) & = & 0.034712 + 0.21210 \,\, \eta/2 ,\\
\lambda_{00} & = & 1.87510,
\end{eqnarray}
obtained for the one-side-clamped and one-side-free vibrating beam. Our calculation takes into account a small static axial (force) load $S$ through  $u <\!\!< 1$, and an end (mass) load term $\eta$ defined as (both have no dimensions):
\begin{eqnarray}
u & = & \frac{S\, h^2}{E_s \, I}, \\
\eta & = & \frac{a l}{w h},
\end{eqnarray}
with $a$ the width of the paddle. 
The sign of $u$ in Eq. (\ref{lambda}) depends on the sign of the load force $S$. With our notations it is $S>0$ for {\it tensile} and $S<0$
for {\it compressive} axial load.
Basically, $u$ compares the axial load $S$ in the beam to the Euler buckling load $E_s I \! /h^2$, while $\eta$ compares the mass of the paddle $\rho_s \, t a l$ to the mass of one foot $\rho_s \, t w h$. 
Note that a tensile axial load {\it brings the frequency down}, as opposed to the 'guitar string', see Appendix \ref{doubleclamped}. 
Note also the numerical difference regarding the influence of axial load on $\omega_0$ with respect to the work of Ref. \cite{mcfarlandsurfacestress} (different $\phi_0(\eta=0)$ coefficient).

When metal is added on the two sides of the structure (with thickness $e_f$ on the front-side, $e_b$ on the back-side and $e_m=e_f+e_b$ the total coating thickness), the problem is equivalent to the one of a device containing 'I-shaped' cantilever beams of moment of inertia $I_a$ and Young modulus $E_s$, total thickness $t_a$ (we now call $e_s$ the silicon thickness instead of $t$) and having a mass density $\rho_a$:
\begin{widetext}
\begin{equation}
\!\!\!\!\!\!\!\!\!\!\!\!\!\!\!\!\!\!\!\!\!\!\!\!\! I  \rightarrow  I_a= 
\frac{e_s^4 w^2 + e_m^4 \tilde{w}^2 + 4 e_s \tilde{w} 
\left[ e_s^2 e_m w + e_m \left(e_f^2 w+e_b^2 w-e_f e_b (w-3 \tilde{w}) \right) + 3/2 \, e_s (e_f^2 w + e_b^2 w + 2 e_f e_b \tilde{w})\right] 
}{12\,(e_s w + e_m \tilde{w})} , \label{Ia}
\end{equation}
\begin{eqnarray}
\rho_s &\!\!\! \rightarrow & \!\! \rho_a= \frac{\rho_s e_s + \rho_m e_m}{e_s + e_m}, \\
t &\!\!\! \rightarrow & \!\! t_a= e_s + e_m.
\end{eqnarray}
\vspace*{2mm}
\end{widetext}
We introduced the renormalized width $\tilde{w}=w \, E_m/E_s$ corresponding to both the front and back part of the fictive 'I-shape', with $w$ the middle width. The metal characteristics are written $\rho_m$ (density) and $E_m$ (Young modulus). With the $x$ axis pointing in the direction of the front-side layer and having its zero in the center of the silicon part, the neutral fiber's position is $x_0=(e_f-e_b)$ $\frac{(e_s+e_m)\tilde{w}}{2(e_s w +e_m\tilde{w})}$.
With $e_b=0$ we recover the 'T-shaped' beam result of a single-sided coating \cite{timo}, and of course the above bare silicon case with $e_b=e_f=0$. Note that the definition of $\eta$ is unchanged. \\

This modeling considers perfectly symmetric and flat structures. However, due to the fabrication process an asymmetry is always present (up to $5-10~$\%) together with a slight thickness gradient along the feet (up to $10-20~$\%). The assymmetry could impact the non-linear (geometric) behavior of the structures, but no conclusive measurements on that point have been obtained yet. The thickness gradient is thought to affect both the spring constant $k$ and the mass $m$ in roughly the same way, leaving the resonance frequency $f_0$ practically unchanged (see \cite{JLTP_VIW} for a treatment based on Rayleigh's method). Therefore, the experimentally deduced parameters quoted in the present article are average values (within typically 10$~$\%), and we believe that the imperfections of the devices do not affect our conclusions.

\subsection{Stored stress}

We now turn to the definition of the static axial (force) load $S$. 
Both the silicon substrate and the metal layers store stresses/strains (see e.g. \cite{nanotribo}). The first obvious origin is the mismatch between their thermal expansion coefficients (bimetallic strip effect). Internal stresses, of intrinsic origin, are also at work. Their physical nature is complex, and depend on the growth process of both the silicon sample and the coating layers. Lattice mismatch is also included in this intrinsic stress.

We use here an adapted version of the model first developed by S. Timoshenko for thermal stresses in a bi-layer composite \cite{thermalTimo,stoneymultilayer}. 
For a well-adhered surface, the stress $\vec{\sigma}$ in the bulk of the material is parallel to the interface. 
The situation at the edges can be rather complex \cite{ZhangZhao1}, but shall not be discussed in the present work. Indeed, 
the major hypothesis of the model is that the strains stored in each layer (metal $\varepsilon_m$ and substrate $\varepsilon_s$) are homogeneous, leading to uniform bending moments and axial loads ($S$ being the one felt by the silicon layer), with discontinuities in the stress tensor at the interfaces.
Furthermore, the strains are assumed to be equal in the front and back layers since the coating material is the same.  
These points are clearly a limitation to the model \cite{yinZhang}, and the values quoted in the present article shall be taken as average parameters. 
Since we are dealing with thin films, we believe that these assumptions should not weaken the conclusions of the article.

The second important assumption of the model is that the radius of curvature is the same for all layers, which is valid for small distortions. 
We call $I_f=1/12 \, w e_f^3$, $I_b=1/12 \, w e_b^3$ and $I_s=1/12 \, w e_s^3$ the moments of inertia calculated for the front-side layer, the back layer and the silicon substrate separately. 
Using the equilibrium condition and the absence of slip at the interface we obtain the relation:
\begin{widetext}
\begin{eqnarray}
S & = & - \, \frac{w \, (\varepsilon_s-\varepsilon_m)}{\frac{1}{E_s^{*} e_s}+\frac{4 E_s^{*} I_s + E_m^{*} \left[ 4 I_f + 4 I_b + w e_f \left( e_f+e_s \right)
\!\!\! 
\left( (e_f+e_s)- \alpha \frac{e_b}{e_f} (e_b+e_s) \right)  \right]}{4 E_m^{*} e_f \left( 1+ \frac{e_b}{e_f} \alpha \right)\left[ E_m^{*} \left(I_f+I_b \right)+ E_s^{*} I_s \right]}},  \label{Svalue}
\end{eqnarray}
with 
\begin{eqnarray}
\alpha & = & \frac{4 E_s^{*} I_s + E_m^{*} \left[ 4 I_f + 4 I_b + w e_f (e_f+e_s) (e_f+e_b+2 e_s) \right] }{4 E_s^{*} I_s + E_m^{*} \left[ 4 I_f + 4 I_b + w e_b (e_b+e_s) (e_f+e_b+2 e_s) \right] }. \label{alphacoeff}
\end{eqnarray}
\vspace*{5 mm}
\end{widetext}
In the expressions above a star has been used to denote the biaxial Young modulus defined as $E_i^{*} = \frac{E_i}{1 - \nu_i}$, with $\nu_i$ the Poisson ratio of material $i$.
For $e_b=0$ one recovers Timoshenko's result (assuming thermal $\varepsilon_i\,$s) \cite{thermalTimo}.
In the thin films case $e_f,e_b <\!\!< e_s$ the expressions write simply 
$\alpha \approx 1$ and $S \approx - E_m^{*} (\varepsilon_s-\varepsilon_m) \, e_m w$; the axial load is proportional to the {\it total amount of metal} deposited.
In Appendix \ref{stoney} we give a summary of the static properties of the beam, namely the forces in each layer, their related moments of flexion and the global distortion (i.e. the Stoney problem adapted to a tri-layer).
According to recent theoretical work, for a bi-layer this type of analytical results is robust for long and thin beams in the linear regime with ideal interfaces \cite{sadercantilever2001,ZhangZhao1,ZhangZhao2}.

In the following experimental part, we will be dealing with the {\it 'total stress' stored in the metallic layers} defined as  $\sigma_m= - S/(e_m w)$.
From the preceding we immediately notice that this parameter will be roughly independent of the geometrical dimensions, and characteristic of the metal-silicon couple under study. 

\begin{figure}[t!]
\includegraphics[height=6.3 cm]{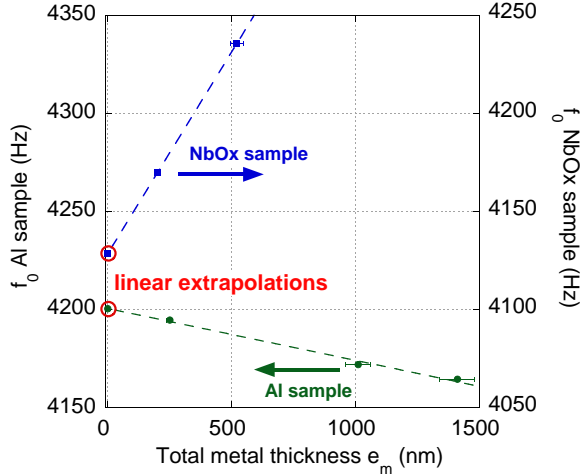}
\caption{\label{freqs} (Color online) T=0$~$K resonance frequencies extracted from the measurements of samples E6 (aluminum coating) and Ec4b (oxidized niobium) as a function of the total amount of metal deposited. Dashed lines are guides to the eyes, showing the $e_m=0$ extrapolation (circled).}
\end{figure}

\section{RESULTS}

We now turn to the interpretation of the parameters $m$, $k$ and $\Lambda_1(x)$ extracted from our data, using the theoretical modeling of the previous section.
\vspace{2mm}

The zero-temperature frequency of the first mechanical mode (as extracted from Fig. \ref{freqvsT}, see below) is plotted in Fig. \ref{freqs} for samples E6 (aluminum) and Ec4b (oxidized niobium) as a function of the total amount of metal characterized by its thickness $e_m$. The metal depositions have been done {\it on both sides} of the samples, and are summarized in Table \ref{evaps}.
\begin{table}[h!]
\begin{center}
\begin{tabular}{|c|c|c|}    \hline
Deposition                       &  E6 (Al)          &     Ec4b (NbOx)     \\    \hline    \hline
 1$^{rst}$ dep., front side      &  $ 250\,$nm       & $ 200\,$nm                    \\    \hline
 2$^{nd}$  dep., back side       &  $+760\,$nm       & $+320\,$nm                    \\    \hline
 3$^{rd}$  dep., front side      &  $+400\,$nm       & X                             \\    \hline
\end{tabular}
\caption{\label{evaps} Characteristics of the metal depositions performed on E6 and Ec4b. Values within typ. $\pm5~$\%.}
\end{center}
\end{table}

\subsection{Stress and elasticity of metal}
\label{valstress}

In Fig. \ref{freqs} the frequency of both resonators is observed to shift smoothly with the total metal thickness $e_m$.
Using the expressions given in Appendix \ref{stoney} and the geometrical characteristics of the devices, one can easily see that for the first and third depositions (Table \ref{evaps}) the bending moments induced by the metals are maximal, while they are close to zero for the second data points. We see no evidence of any influence of this bending moment. We conclude that as far as stresses are concerned, only the axial force load $S$ influences the frequency.
This point is usually an assumption in the literature \cite{mcfarlandsurfacestress}.

In order to keep the discussion as simple as possible, we take for all materials an average Poisson ratio of $\nu \approx 0.3$. We checked that this choice does not affect the spirit of our conclusions.

From equations (\ref{omegadef}) and (\ref{kspring})-(\ref{alphacoeff}) giving the expressions enabling the calculation of $f_0$, we see that three metal parameters have to be taken into account to fit quantitatively Fig. \ref{freqs}: the mass density $\rho_m$, the stress $\sigma_m$ and the Young modulus $E_m$. For the oxidized niobium samples, the metal density is increased by about 1.7$~$\%. 

 Although the mass densities $\rho_m$ are well known (better than $1~$\%), the exact values of the stored stresses and the Young moduli at low temperatures are difficult to access. We will therefore compute the frequencies of the structures for a series of couples $(\sigma_m,E_m)$. Note that depending on the rigidity, hardness and density of the metal, the frequency can increase (niobium) or decrease (aluminum) with the metal quantity.
\begin{table}[h!]
\begin{center}
\begin{tabular}{|c|c|c|}    \hline
Parameter                 &  E6                &     Ec4b            \\    \hline    \hline
 $h$                      &  $ 1.35\,$mm       & $ 1.12\,$mm                   \\    \hline
 $w$                      &  $32.5\,\mu$m      & $15.2\,\mu$m                  \\    \hline
 $e_s$                    &  $9.5\,\mu$m       & $7.0\,\mu$m                   \\    \hline \hline
 $l$                      &  $1.55\,$mm        & $1.30\,$mm                    \\    \hline
 $a$                      &  $27.0\,\mu$m      & $16.0\,\mu$m                  \\    \hline
\end{tabular}
\caption{\label{geom} Geometrical characteristics of the bare silicon beams of E6 and Ec4b. Note that these values are average parameters neglecting asymmetries and thickness gradients.}
\end{center}
\end{table}

Of course the main parameters defining the resonance frequency are those of the silicon. 
We take for the T=0$~$K Young modulus of silicon in the $<110>$ direction the value of $E_s=170.2~$GPa, in close accordance with \cite{siliconE} (1$~$\%). Since at first order the frequency depends linearly on $e_m$, we extrapolate the data to $e_m=0$ to extract the bare silicon resonance frequency, see Fig. \ref{freqs}. From SEM images, we know the geometry of the silicon beams within $5~$\% approximately. We then adjust carefully the dimensions within these error bars to fit the bare resonance frequencies. Results are summarized in Table \ref{geom}. These parameters are then injected in the theory, Sec. \ref{theor}, equations (\ref{kspring})-(\ref{alphacoeff}), where only $(\sigma_m,E_m)$ are unknown.

\begin{figure}[t!]
\includegraphics[height=6.5 cm]{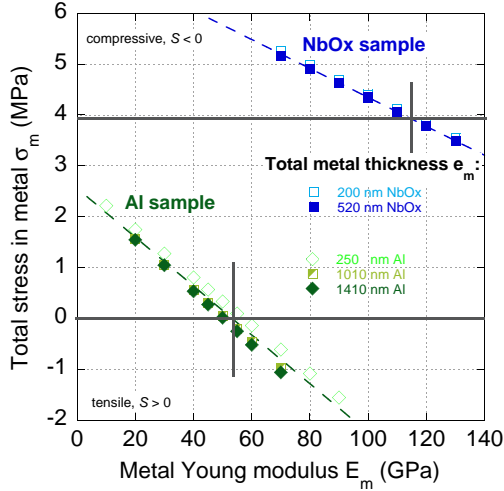}
\caption{\label{stressandYoung} (Color online) T=0$~$K couples $(\sigma_m,E_m)$ fitting the resonance frequencies of Fig. \ref{freqs} for samples E6 (aluminum) and Ec4b (oxidized niobium). 
Light colors correspond to the thinnest layers while dark colors represent the thickest.
The dashed lines are guides to the eyes, showing the very small dispersion of the data (about $\pm5~$\% for the aluminum sample and $\pm2~$\% for the oxidized niobium sample, on both axis). The full lines represent the retained couples.}
\end{figure}

In Fig. \ref{stressandYoung} we give the resulting couples $(\sigma_m,E_m)$ fitting the two samples T=0$~$K resonance frequencies. Each point is obtained by fixing a Young modulus and adjusting the corresponding stress $\sigma_m$. All data collapse on single lines, within about $\pm5~$\% for the aluminum sample and $\pm2~$\% for the oxidized niobium sample, on both axis.
This very small dispersion demonstrates the validity of the model developed in Sec. \ref{theor}. The remaining discrepancies are believed to be due to both errors on our experimental parameters (about $5~$\%) and limitations of the model (for instance, a slight thickness-dependence of the total metallic stress $\sigma_m$).

At room temperature, the single-sided oxidized niobium samples are clearly bent under a {\it tensile stress}, Fig. \ref{pictures}. This curvature is not visible anymore when both sides are coated with metal, in accordance with Appendix \ref{stoney}. 
From Stoney's formula and the deflections measured with SEM images we obtain metal stresses $\left|\sigma_m (300~K) \right|$ around 300$~$MPa \cite{stressbend}, which is two orders of magnitude larger than the stress observed in Fig. \ref{stressandYoung} at 4$~$K. 
On the other hand, no curvature could be seen at room temperature on samples having aluminum coatings, regardless of the thickness evaporated.
We thus consider that for aluminum $\sigma_m (300~K) \approx 0~$MPa.

We estimate the thermal stresses generated while cooling from 300$~$K to 4$~$K from the thermal expansion coefficients (bimetallic strip effect) \cite{thermexpSi,thermexpSiTheror,thermexpAl,thermexpNb,thermalexpAll}.
We obtain {\it compressive stresses} around $\left|\Delta \sigma_m \right| \approx 250-300~$MPa for all samples (aluminum and oxidized niobium). Again these values are far larger than the observed ones at 4$~$K.  

As a matter of fact, from the two preceding paragraphs dealing with $\sigma_m$ we can interpret Fig. \ref{stressandYoung}:
\begin{itemize}
\item Our oxidized niobium films store stresses in the range $3-5~$MPa, for Young moduli in the range $60-120~$GPa. 
The (interstitial) oxygen doping increases the room temperature niobium Young modulus $E_m$ by typically 4$~$\% \cite{dopingnb}.
From the niobium literature \cite{niobE,fermilab}, pure polycrystalline niobium exhibits a Young modulus at low temperature about 7$~$\% higher than the 300$~$K value. We therefore expect in our experiments the oxidized niobium $E_m$ to be around 115$~$GPa, which corresponds for our samples to a stress $\sigma_m$ of 3.95$~$MPa (full lines Fig. \ref{stressandYoung}).
We conclude that the initial stored stress is {\it practically compensated} at 4$~$K by thermally-induced stress.  \\
The (lowest) yield strengths quoted at room temperature for untreated niobium are around $100-200~$MPa, and can be quite larger at 4$~$K \cite{fermilab}, up to $5-6$ times higher. Furthermore, the oxygen doping increases substantially the hardness of the metal at room temperature (in our case, it should be about an order of magnitude) \cite{nboxsupra}. We conclude that our oxidized niobium films should exhibit at cryogenic temperatures yield strengths easily above $500~$MPa.
\item It is clear that our aluminum sample {\it does not display} total metal stresses in the 100$~$MPa range, which means that the thermal stress has been released by some means. 
The hardness and rigidity of a metal are directly dependent on its microscopic structure (grains, voids, microcracks, dislocations, etc...) \cite{grains}. In particular, {\it anelastic effects} due to grain boundary sliding can result in a reduced effective Young modulus with respect to the 'bulk value' \cite{anelAlu}. \\
\begin{figure*}[t!]
\includegraphics[height=7. cm]{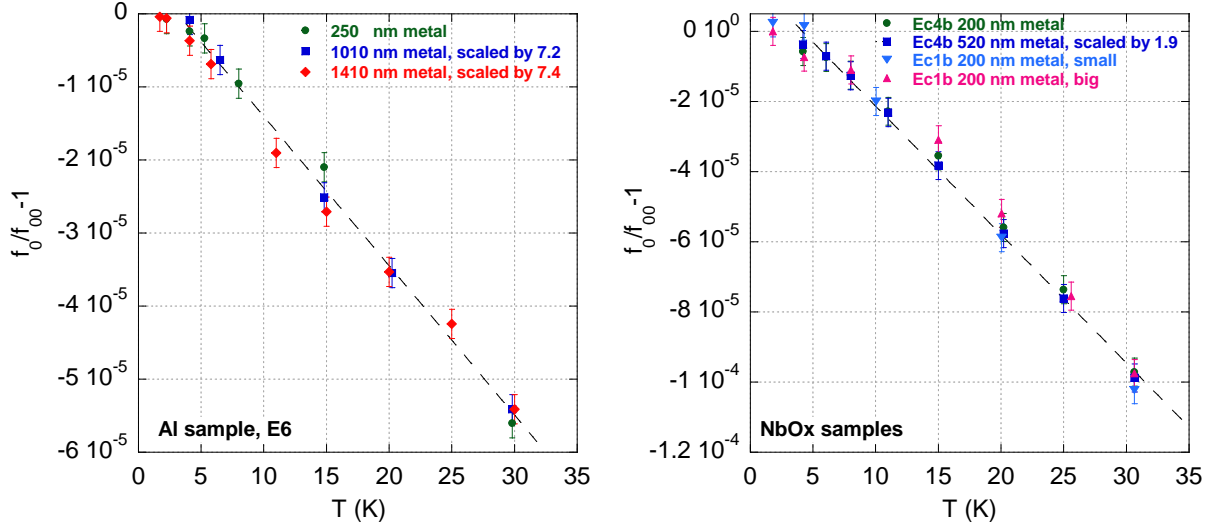}
\caption{\label{freqvsT} (Color online) Normalized resonance frequency shifts $f_0(T)/f_0(T=0)-1$ of our structures as a function of temperature. Left: results on E6, aluminum-coated. The three evaporations are displayed, scaled on each other. Right: results for Ec1b (double-structure) and Ec4b (multiple-depositions), with oxidized niobium coatings. {\it Only} the second niobium deposition has been scaled for the graph. Error bars are about $\pm 10 ~$mHz. The dashed lines are guides to the eyes. From these data we extract the T=0$~$K frequencies plotted in Fig. \ref{freqs}.}
\end{figure*}
The structure of a thin film is directly related to the deposition technique \cite{hanbookthinfilm}.
Sputtered aluminum thin films exhibit at room temperature mechanical properties close to the bulk material, or even better \cite{alSputter, alSputter2}.
However, it is known that Joule-evaporated aluminum films can be quite soft \cite{softalu}: the Young modulus is smaller than its bulk value, and the elastic limit is low.  We expect $E_m$ in the $45-75~$GPa range \cite{anelAlu}, which means for our sample a 4$~$K stored stress around zero. We therefore retain the couple $E_m \approx 55~$GPa and $\sigma_m \approx 0~$MPa for our experiments (completely relaxed stress, full lines in Fig. \ref{stressandYoung}). \\
Furthermore, in order to be consistent with this absence of stress, we expect an elastic limit for our aluminum coatings as low as $10-50~$MPa \cite{softalu}.
\end{itemize}
These properties will be invoked again in the discussion on the {\it anelasticity} of the metallic thin films.

\subsection{Temperature dependencies}
\label{sectemp}

We first discuss the position $f_0(T)$ of our measured resonance peaks.
In Fig. \ref{freqvsT}, we present the temperature dependence of the first mechanical mode resonance frequencies of our structures. What is plotted is the normalized frequency shifts $f_0(T)/f_0(T=0)-1$. {\it Only} the multiple-deposition data have been scaled in order to display a single curve (see legend). The zero temperature frequencies in Fig. \ref{freqs} are obtained through the extrapolation of these data.

The two panels in Fig. \ref{freqvsT} (aluminum / oxidized niobium) share identical characteristics:
\begin{itemize}
\item The dashed lines, which are guides for the eyes suggest {\it linear} temperature dependencies down to 4$~$K with a flattening below. All the frequencies {\it decrease} with increasing temperature.
\item The slope of the linear dependence depends on the metal quantity, and {\it increases} with the amount of metal (see scaling factors). Aluminum and niobium display shifts of the same order.
\item Moreover, for the double-structure Ec1b the 'big' and 'small' oscillators display the {\it same} normalized shifts. They share the same metal (oxidized niobium, 200$~$nm), and about the same thickness (6.4$~\mu$m, $\pm 0.2~\mu$m). However, the other geometrical dimensions are quite different, and their resonance frequencies are clearly distinct: 1907.35$~$Hz for the 'big' structure, and 3248.45$~$Hz for the 'small', at T=0$~$K.
\end{itemize}
From the theory in Sec. \ref{theor}, we easily realize that if the metal stress term $\sigma_m$ would be temperature-dependent (being the same for both the 'big' and the 'small' structures), we should observe different normalized frequency shifts for the two Ec1b oscillators (Eq. (\ref{lambda}) and related ones).
Moreover, below typically 40$~$K, the thermal expansion coefficient is constant \cite{thermalexpAll} for silicon \cite{thermexpSi,thermexpSiTheror}, aluminum \cite{thermexpAl} and niobium \cite{thermexpNb}. We thus have to conclude that the temperature-dependence seen in Fig. \ref{freqvsT} {\it is not due} to a temperature-dependent (thermal) stress in the beams: the stress is constant over the range $1.5-35~$K, and we should seek another explanation to Fig. \ref{freqvsT}.

The dimensions being temperature-independent, and postponing the discussion on the reactive component of the friction mechanisms, the only other source of 'softening' lies in the materials' elasticity. 
The increase of the effect seen in Fig. \ref{freqvsT} with the amount of metal deposited is {\it inconsistent} with a temperature dependence of the silicon Young modulus $E_s$. Below typically $50~$K, $E_s$ flattens out \cite{siliconE}; here we have to take it as constant at the level of  $10^{-5}$.

\begin{table}[h!]
\begin{center}
\hspace*{-3 mm}
\begin{tabular}{|c|c|c|c|c|}    \hline
Deposition                &  E6  (Al)          &  E6  (Al)         &     Ec4b (NbOx)            &     Ec4b (NbOx)         \\    
                          &  freq.             &  width            &      freq.                 &      width              \\ \hline    \hline
   T=0$~$K val.           &  $ 4194.65\,$Hz    & $ 13.5\,$mHz      &  $4169.85\,$Hz             &   $97.5\,$mHz           \\  
  1$^{rst}$ dep.          &   1                &      1            &        1                   &        1                \\    \hline
  2$^{nd}$  dep.          &  7.20              &     7.15          &       1.90                 &        1.80             \\    \hline
  3$^{rd}$  dep.          &  7.40              &     7.55          &        X                   &       X                 \\    \hline 
\end{tabular}
\caption{\label{scalings} Scaling factors for samples E6 and Ec4b, as a function of the metal deposition. The T=0$~$K values are also quoted.}
\end{center}
\end{table}

Although the linear dependence of $f_0$ seen in Fig. \ref{freqvsT} is very weak, it is thus a signature of the temperature dependence of the {\it metallic coatings}' Young moduli $E_m(T)$ \cite{niobE,fermilab,aluE,theorE}. 
We write $E_m (T)=E_m(0)+ \delta E_m(T)$. Taking a first order expansion of Eq. (\ref{Ia}) in $\delta E_m(T)/E_s$, and $e_m/e_s$ (thin film limit) we obtain:
\begin{equation}
\frac{f_0(T)}{f_0(T=0)} -1 = \frac{3}{2} \, \frac{\delta E_m(T)}{E_s}  \frac{e_m}{e_s} .
\end{equation}
This relation explains why the effect increases (close to linearly) with the metal quantity, and why we observe the same normalized shifts for the two oscillators of structure Ec1b. The observed normalized shift is not rigorously proportional to the metal thickness, certainly because of the thickness-dependence of the materials parameters (which has been neglected in the present work).
We typically obtain (linear) changes from 0$~$K to 35$~$K of the order of  $-0.5~$\% for both aluminum and (oxidized) niobium Young moduli, consistent with the literature. 
In terms of resolution, the $\pm10~$mHz error bars of Fig. \ref{freqvsT} correspond to a relative error on $E_m$ of $10^{-4}$, while the absolute precision is only related to our knowledge of the devices characteristics, limited in our case by the geometry with typically $\pm5~$\%.  

\begin{figure*}[t!]
\includegraphics[height=7. cm]{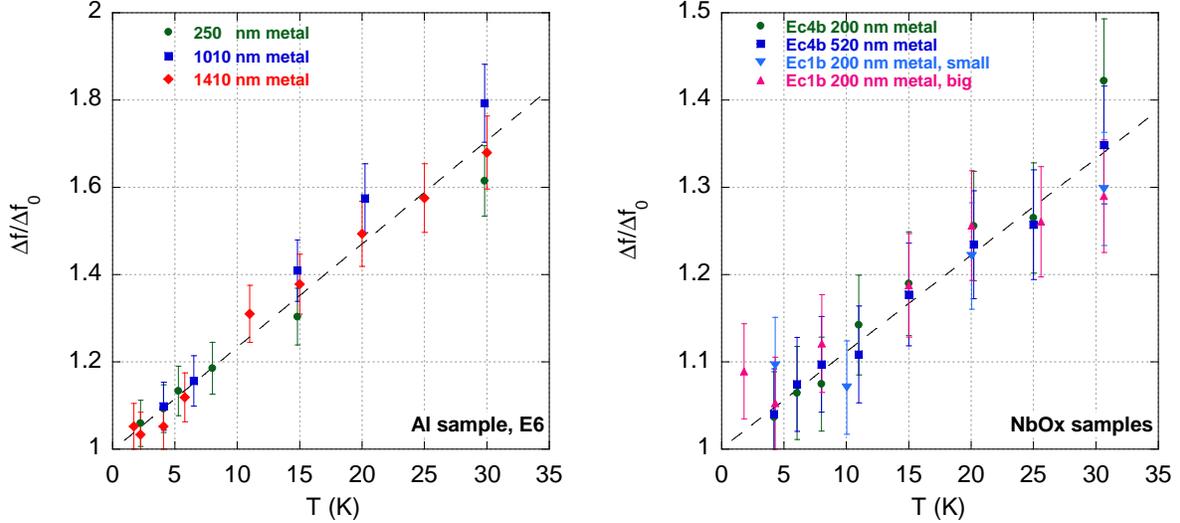}
\caption{\label{widthvsT} (Color online) Normalized linewidths $\Delta f(T)/\Delta  f(T=0)$ of the mechanical resonances of our structures as a function of temperature. Left: results on E6, aluminum-coated. Right: results for Ec1b (double-structure) and Ec4b (multiple-depositions), with oxidized niobium coatings. {\it No scalings} have been applied for the graph. Error bars are about $\pm 5 ~$\%. The dashed lines are guides to the eyes.}
\end{figure*}

We now turn to the dissipation term $\Delta f(T)$. It is clear from Ref. \cite{parpiadissipeMEMS} that the quality factor $Q$ of our devices is limited by the friction occurring in the metallic coatings.
In Fig. \ref{widthvsT} we present the linewidths measured for our first mode mechanical resonances as a function of temperature. What is displayed is the normalized linewidth $\Delta f(T)/\Delta  f(T=0)$. As a result, all data fall on two universal curves characteristic of each metal. No scalings have been applied, and the metal thickness dependence is directly incorporated in $\Delta  f(T=0)$. The two metals display the same tendency: the damping measured through $\Delta f$ {\it increases} with increasing temperature almost linearly.
 
We define a linewidth scaling factor by taking the ratio of the $\Delta  f(T=0)$ measured for various metal quantities to the one measured for the first metal deposition.
This scaling term together with the scaling factor of the frequency shifts in Fig. \ref{freqvsT} are summarized in
 table \ref{scalings}, for samples E6 and Ec4b.
From Table \ref{scalings} we realize that the scalings of the frequency shifts and the linewidths {\it are the same}. Moreover, the dashed guide to the eyes in Fig. \ref{widthvsT} suggests a linear temperature dependence, as in Fig. \ref{freqvsT}. We conclude that the two effects, namely the frequency shift and the dissipation, {\it originate in the same mechanisms}, linked to the metal Young moduli $E_m$.

This result alone is not surprising. Indeed, since the acoustic susceptibility links frequency shift (i.e. $\Lambda_2$) and linewidth (i.e. $\Lambda_1$) through the Kramers-Kronig relations, for a given friction mechanism, they should both scale in the same way.
However, although the 'small' and 'big' structures of sample Ec1b share the {\it same} normalized shift, their zero-temperature linewidth $\Delta f (T=0)$ are {\it very different} (see Table \ref{dissipe} below). 
We thus have to conclude that in our case they {\it cannot be} the two signatures of the same friction mechanism.
Consequently the reactive component of the friction has to be negligible for our devices (i.e. $\Lambda_2$ too small to be measured), while the frequency shift is solely due to the metals softening with increasing temperature.

\subsection{Dissipation parameters}

Understanding dissipation mechanisms in micro and nano mechanical objects at low temperatures is an issue, for both device optimization and fundamental physical understanding \cite{parpiadissipeMEMS,gravit,thermodamp,glassTLS,castroNeto,nanodissipe,clamploss}. 
From Fig. \ref{widthvsT} we extract a zero temperature linewidth $\Delta  f(T=0)$ for each structures' resonance. This value is converted into a zero temperature damping term $\Lambda_1 (T=0) = m \, \pi \Delta f(T=0)$. 
In table \ref{dissipe} we summarize these zero-temperature dissipation parameters for all our samples with a single layer deposited, together with other relevant values. Clearly, the damping depends on both the {\it coating material} and the {\it geometry} of the oscillator.

\begin{table}[h!]
\begin{center}
\hspace*{-3 mm}
\begin{tabular}{|c|c|c|c|c|}    \hline
Parameter                 &  E6                &  Ec4b                 &     Ec1b 'small'       &     Ec1b 'big'            \\     
                          &      (Al)          &      (NbOx)           &        (NbOx)          &       (NbOx)              \\    \hline    \hline  
   $\Delta f(T=0)$        &  $ 13.5\,$mHz      & $ 97.5\,$mHz          &    $82\,$mHz           &       $38\,$mHz           \\    \hline
   $\Lambda  $            &   $56~$pN.s/m      & $153~$pN.s/m          &    $93~$pN.s/m         & $585~$pN.s/m       \\    \hline    \hline
   $f_0(T=0)$             &   $4194.65~$Hz     & $4169.85~$Hz          &    $3248.45$Hz         & $1907.35~$Hz       \\    \hline
   $Q$                    &   $0.31\,10^6$     &  $43\,10^3$           &      $40\,10^3$        &  $50\,10^3$        \\    \hline \hline
   $k$                    &    $0.95~$N/m      &       $0.34~$N/m      &       $0.15~$N/m       &     $0.70~$N/m     \\    \hline
  $m$                     &    $1.35~\mu$g     &       $0.50~\mu$g     &        $0.36~\mu$g   &        $4.90~\mu$g   \\    \hline
  $w$                     &    $32.5~\mu$m     &       $15.2~\mu$m     &        $14~\mu$m     &        $120~\mu$m    \\    \hline 
  $h$                     &    $1.35~$mm       &       $1.12~$mm       &        $1.25~$mm     &        $1.65~$mm     \\    \hline \hline
  $e_s$                   &    $9.5~\mu$m      &       $7.0~\mu$m      &        $6.2~\mu$m    &        $6.65~\mu$m   \\    \hline 
  $e_m$                   &    $250~$nm        &       $200~$nm        &        $200~$nm      &        $200~$nm      \\    \hline 
\end{tabular}
\caption{\label{dissipe} Zero temperature dissipation obtained for the first deposition coated devices. Other parameters relevant to the discussion are also given.}
\end{center}
\end{table}

Extrinsic noise, mechanical heating, and Joule heating can all be disregarded \cite{JLTP_VIW}.
From Sec. \ref{sectemp} we know that the damping originates in the Young moduli of the metallic coatings $E_m$. It is thus natural to invoke {\it thermoelastic damping}, the process by which the acoustic waves generated by the motion of the beam decay into heat \cite{thermodamp}. Following these authors, we write:
\begin{eqnarray}
\frac{f(T)}{f(T=0)} - 1 & = & \Delta_E(T) \, g_2[\xi(T)] ,\\
Q^{-1}(T) &= & \Delta_E(T) \, g_1[\xi(T)],
\end{eqnarray}
with $g_1(\xi)$ and $g_2(\xi)$ two functions bounded between $[0,\frac{1}{2}]$. The temperature dependencies have been written explicitly.
$\Delta_E$ is a material-dependent parameter given by $\Delta_E=E_i \, \alpha_i^2 \, T/C_{p,i}$ ($\alpha_i$ the thermal expansion coefficient and $C_{p,i}$ the specific heat of material $i$), and $\xi=(\pi/\sqrt{2}) \sqrt{\omega_0 \tau}$ compares the motion frequency $\omega_0$ to the thermal relaxation time $\tau$ of the mechanical mode. The geometry enters the model only through the thickness $t$ of the beam with $\tau=t^2/(\pi^2 \chi_i)$ (and $\chi_i$ the thermal diffusivity of material $i$).

The model described in \cite{thermodamp} considers a beam made of a single material. In the following order of magnitude discussion, we will separately consider the silicon substrate and the metallic coatings. We evaluate the two parameters $\Delta_E$ and $\xi$ at 30$~$K, knowing that they will both decrease for lower temperatures. \\
For monocrystalline silicon \cite{thermodamp} one obtains for $\Delta_E$ about $5. 10^{-8}$ with $\xi \approx 10^{-3}$, 
confirming that the substrate can be safely neglected. \\
For aluminum at the same temperature, the number comes out to be around  $\Delta_E \approx 10^{-5}$ with $\xi \approx 10^{-3}$, and for (oxidized) niobium we obtain also $\Delta_E \approx 10^{-5}$ with $\xi \approx 5. 10^{-4}$ (the films have a low Residual Resistive Ratio). In the regime $\xi <\!\!<1$, the functions $g_1,g_2$ can be approximated by :
\begin{eqnarray}
g_2[\xi] & = & \frac{17}{840} \xi^4, \\
g_1[\xi] & = & \frac{1}{5} \xi^2.
\end{eqnarray}
The numbers we obtain are clearly too small, and the model fails to describe our results: 
in Table \ref{scalings} the damping is close to proportional to $e_m$ while in Table \ref{dissipe} we obtain a width $w$ or length $h$ dependence as well, which cannot be explained here. The close to linear temperature dependence in Fig. \ref{widthvsT} is extremely different from the power law decrease of $\alpha_i$.

\begin{figure*}
\includegraphics[height=7cm]{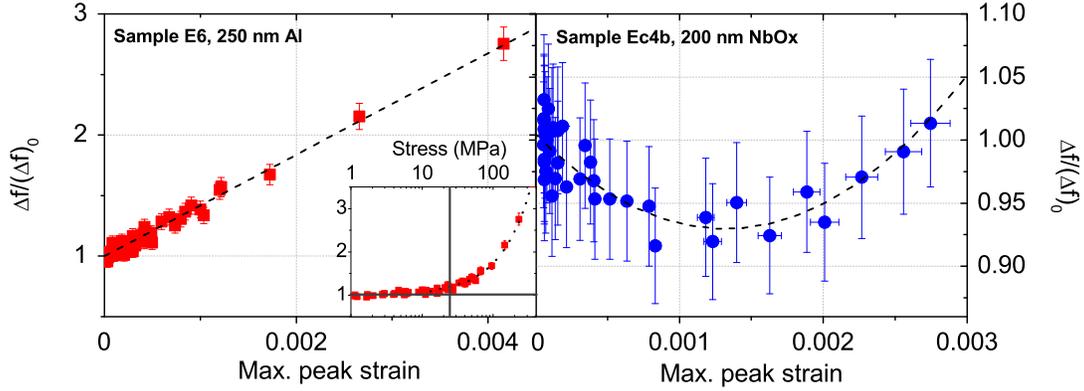}
\caption{\label{anelast} (Color online) Anelasticity of the metallic coatings at 4$~$K. The plot displays the linewidth of the resonance normalized to the zero strain linewidth, as a function of the maximum peak strain. Left: aluminum sample E6 at first metal deposition. The inset shows the same data in a log-lin. plot as a function of the maximal peak stress in the metal, zooming on the onset of the damping increase (marked by full lines). Right: oxidized niobium sample Ec4b, first metal deposition. The dashed lines are guides emphasizing the two very different behaviors measured.}
\end{figure*}

The second obvious mechanism which has to be discussed is {\it clamping losses}, the process by which the acoustic waves generated by the motion of the beam radiate away into the supporting substrate \cite{clamploss}.
For all our structures, the relevant mode wavelength is of the order $\lambda_{rad} \approx 4.3 \, h$ ($\pm 0.2\,h$). We are clearly in the regime $\lambda_{rad} >\!\!> t,w$ (thin beam) and $\lambda_{rad} >\!\!> t_{sup}\approx 300~\mu$m (thin support). 
From \cite{clamploss} we write, neglecting the composite nature of our beams:
\begin{equation}
Q^{-1} \approx 0.95 \frac{w}{h} \frac{t^2}{t_{sup}^2} .
\end{equation}
The values obtained are around $10^{-5}$ which is of the right order of magnitude (within a factor 10). However, this expression is unable too to describe the geometry dependence of Table \ref{dissipe}.
Moreover, note that samples Ec1b 'small' and Ec4b which are very similar in dimensions and resonant frequency, have different clamping geometries (see Fig. \ref{pictures}), but practically identical dampings (Table \ref{dissipe}).  

From the preceding we have to conclude that the damping is linked directly to the dynamics occurring in the metal layer, but the exact mechanism is not known.
From Table \ref{dissipe}, we realize that the oxidized niobium coated samples have a damping which clearly depends on the geometry. Looking carefully at the data, it seems to depend on the length $h$ and scales as $1/h^{n}$ ($n$ around 2.5).
Altogether, we thus have to write $\Delta f =C\!ste \, (e_m/e_s)(E_m/E_s)/h^{n}$ which fits our whole set of experimental results to about $\pm 2~$\% (the constant $C\!ste$ being material-dependent).
 We can speculate that the friction mechanism responsible for the damping in both metals is also responsible for the anelasticity discussed below in aluminum samples, namely grain boundary sliding \cite{anelAlu}.

\subsection{Anelasticity}

The technique used in the present work enables very large distortions of the structures to be achieved. We can thus study the dynamics of the metallic films deposited under very high (dynamic) strains. From \cite{JLTP_VIW} we write the {\it ac} contribution to the stress tensor in the silicon at the interface with the front metallic layer:
\begin{equation}
\delta \sigma_s = \frac{6}{e_s w} \frac{h}{e_s} \left( 1- \frac{z}{h} \right)\, (k\, x). \label{stress}
\end{equation}
The stress is in-plane, and $z$ denotes the coordinate along the foot of the structure, running from 0 to $h$. The maximum is of course obtained at the clamping end ($z=0$). Dividing then Eq. (\ref{stress}) by $E_s$, we obtain the strain at the clamp. Since there is no slippage at the interface, the stress in the metal layers is simply deduced using the values of $E_m$ obtained in the preceding section \ref{valstress}.

The dampings of the first mechanical mode of samples E6 (aluminum) and Ec4b (oxidized niobium) are displayed in Fig. \ref{anelast} as a function of this clamping strain, calculated for the peak values of the displacement $x_{max}$. What is plotted is the resonance linewidth $\Delta f$ normalized to its zero strain value, for the 4$~$K data and first metal deposition. 

We obtain two very different behaviors of $\Lambda_1(x_{max})= m \, \pi \Delta f(x_{max})$ for aluminum and oxidized niobium, characteristic of the damping mechanisms at work. The aluminum layer displays a very strong non-linear dissipation, with seemingly a linear increase with strain (dashed lines in Fig. \ref{anelast}, left panel). On the other hand, the oxidized niobium layer has a practically constant dissipation term (within $\pm 5~$\%), with even a small minimum.

These results have to be correlated with the discussion of Sec. \ref{valstress}. The aluminum layer is soft, with a very low plastic limit. Indeed in the inset of Fig.  \ref{anelast} we plot on a log-lin. scale the normalized linewidth for sample E6, first deposition, as a function of the maximum peak stress in the metal layer. The increase in the damping starts to be visible around about $20-30~$MPa, which suggests a plastic limit around the same value (full lines, inset Fig. \ref{anelast}). On the other hand, for niobium the yield strength is supposed to be above 500$~$MPa, and even for the larger displacements achieved in our studies we never exceed so high dynamic stresses in the metal. Also, we saw for sample E6 (aluminum) in \cite{JLTP_VIW} {\it permanent shifts} after a return to the linear regime from very high displacements, suggesting {\it hardening} of the aluminum $E_m$. This effect was even stronger for thicker metallic layers. Nothing of that sort could be detected with the oxidized niobium layers.

Although there was no measurable non-linear behavior of the spring constant $k(x_{max})$ for both Ec4b and E6 with the first metallic coating, in \cite{JLTP_VIW} we could identify an effect for thicker aluminum layers.
This confirms the absence of a metal friction-induced reactive term at least for very thin films (Sec. \ref{sectemp}). It also demonstrates, for thicker films, the presence of a non-linear Young modulus which is another expression of anelasticity.

The origin of the anelastic behavior is linked to the dynamics occurring in the metal layer, as for the (linear) dampings, but its exact nature is not known. We can speculate that grain boundary sliding is responsible, as it has been suggested for other experiments on soft aluminum films \cite{anelAlu}.
The difference between (oxidized) niobium and aluminum would then be due to the very different grain structure of the layers (Fig. \ref{surface}).

The characteristics of the non-linear dampings discussed here are the same when the amount of metal is increased, with even stronger signatures. Note that the slope of the damping versus strain observed for aluminum in Fig. \ref{anelast} is {\it temperature-independent} \cite{JLTP_VIW}, while the small minimum observed for oxidized niobium decreases with increasing temperature, and disappears around 30$~$K. The aluminum plasticity would provide the stress-relaxing mechanism invoked in Sec. \ref{valstress}.

\section{CONCLUSIONS}

We report on mechanical measurements of cantilever-based silicon MEMS at temperatures between 1.5$~$K and 35$~$K. A (normal-state) metallic layer is used for drive and detection of the motion (magnetomotive scheme), and the aim of the present article is to understand the impact of the coatings on the dynamics, from linear to highly non-linear regime.
Two metals have been studied, namely (soft) Joule-evaporated aluminum and (hard) magnetron-deposited oxidized niobium (containing about 10$~$\% at. O). Multiple metal depositions have been performed, using the same metal, on the same sample.
We benefit from the cryogenic condition to use magnetic fields up to 350$~$mT, and a vacuum below $10^{-6}~$mbar.
Moreover, the temperature is a control parameter together with the driving force, used to characterize thoroughly the oscillators. No anomalous magnetic field $B$ or current $I_0$ dependence could be detected, and only the force $F_0 \propto I_0 B$ has to be considered.

Using especially derived analytical expressions, we extract the influence of the additional mass, additional elasticity and stress due to the metal on the first mechanical mode resonance. 
These expressions are given as tools for data analysis of cantilever MEMS harmonic-drive response.
We demonstrate that only the axial load is relevant, the induced bending moment does not affect the resonance.
We find that for niobium, the initial tensile stress is practically compensated by the compressive thermal stress, leaving only about $+4~$MPa in the metal at low temperatures. For soft aluminum, the stress is relaxed at 4$~$K ($\sigma_m \approx 0~$MPa).
These stresses are found to be temperature-independent in the range 1.5$~$K - 35$~$K.
The mechanical resonance shifts almost linearly with temperature, and the shift corresponds to the softening of the metal Young modulus $E_m(T)$ with the temperature increase. The technique is accurate, with a relative precision reached of $10^{-4}$ and an absolute error defined only by our knowledge of the MEMS characteristics (here about $\pm 5$\%).

We demonstrate that the damping is also related to the metal Young modulus. However, for thin films no reactive component of the friction mechanism is seen. The dissipation (i.e. the linewidth $\Delta f$ of the resonance) cannot be explained by standard models of thermoelastic damping and clamping losses. We thus have to invoke some other friction mechanisms originating in the metallic coatings, for instance grain boundary sliding. The mechanism should explain the peculiar dependence of the linewidth 
 $\Delta f \propto E_m (e_m / e_s)/h^{n}$, with $n$ around $2.5$ and $h$ the length of the feet, $e_s$ the silicon thickness and $e_m$ the total metal thickness.

Furthermore, soft aluminum displays a strong anelastic behavior, as opposed to oxidized niobium, with a seemingly linear dependence of the damping to the (dynamic) strain.
Anelasticity would be the stress-relaxing mechanism invoked for the aluminum coating. As for the damping term, the nature of the mechanism is not known, but we can speculate grain boundary sliding again, presented in the literature as the strongest anelastic phenomenon.
\vspace{2mm}

These results are calling for a precise microscopic understanding, and demonstrate the capabilities of the technique for thin films mechanical characterization.

\appendix

\section{HARMONIC TREATMENT OF DOUBLY CLAMPED BEAM}
\label{doubleclamped}

Following the technique described in \cite{JLTP_VIW}, we give below the expressions describing the case of the doubly clamped beam (length $h$, width $w$ and thickness $t$). The spring constant and the mass write:
\begin{eqnarray}
k & = &   \left(\frac{E_s \, I}{h^3} \right)  \, \lambda^4 \, \left[  \gamma_{0}  \right], \\
m & = &  (\rho_s \, t w h) \left[ \gamma_{0}  \right],  
\end{eqnarray}
with $I=w t^3/12$ and the parameters describing the first mechanical mode:
\begin{eqnarray}
\lambda(u) & = & \lambda_0 \left( 1 + \frac{u}{2} \, \phi_0 \right), \label{beam}\\
\gamma_0  & = & 0.396478 , \\
\phi_0 & = & 0.012289 ,  \\
\lambda_{0} & = & 4.730041 ,
\end{eqnarray}
valid for a stress factor $u <\!\!< 1$ as defined in Sec. \ref{theor}. The sign in Eq. (\ref{beam}) in front of $u$ makes the frequency increase for a {\it tensile} stress $S$, and we recover the expression of \cite{roukesnanotube} as we should.

\section{STATIC PROPERTIES OF TRILAYER}
\label{stoney}

In the following, we briefly summarize the static parameters obtained for the tri-layer adaptation of S. Timoshenko's stressed beam model \cite{thermalTimo,stoneymultilayer}. The assumptions have been given in Sec. \ref{theor}, and we define $F_f$ the total force on the front-side layer and $F_b$ on the back ($S$ being the force acting on the silicon). The corresponding bending moments are $M_f$, $M_b$ and $M_s$ respectively. We have:
\begin{eqnarray}
M_f & = & \frac{E_m^{*} I_f}{r}, \\
M_b & = & \frac{E_m^{*} I_b}{r}, \\
M_s & = & \frac{E_s^{*} I_s}{r}, 
\end{eqnarray}
with $r$ the radius of curvature of the beam and: 
\begin{eqnarray}
F_f & = & -\, \frac{S}{1+\frac{e_b}{e_f} \alpha}, \\
F_b & = & -\, \frac{e_b}{e_f} \alpha \frac{S}{1+\frac{e_b}{e_f} \alpha}, \\
S & = & - (F_f+F_b) ,
\end{eqnarray}
with $\alpha$ and the other terms defined in Sec. \ref{theor}. The force $S$ exerted on the silicon layer is given by Eq. (\ref{Svalue}). The curvature $r$ is then obtained as:
\begin{equation}
\frac{1}{r} = -\, \frac{S}{ \left( 1+\frac{e_b}{e_f} \alpha \right) }\, \frac{(e_f+e_s)- \frac{e_b}{e_f} \alpha (e_b+e_s) }{2 \left( E_m^{*} I_f + E_m^{*} I_b + E_s^{*} I_s \right)}. \label{last}
\end{equation}
Calling the shape of the beam $\upsilon(x)$, from Eq. (\ref{last}) and $1/r(x)=d^2 \upsilon(x)/dx^2$ one calculates easily the distortion of the cantilever ($\upsilon(0)=0$, $d \upsilon(0) / d x =0$). 
In particular, for a single thin film over a thick substrate ($e_f <\!\!< e_s$, $e_b=0$) one recovers Stoney's result \cite{purestoney}.
Note that for thin films the axial load $S$ {\it increases roughly as} $e_m=e_f+e_b$ while {\it the bending moments cancel if} $e_f=e_b$.

\begin{acknowledgments}
We wish to acknowledge the support of Thierry Fournier, Christophe Lemonias, and Bruno Fernandez in the fabrication of samples. 
The authors wish to acknowledge valuable discussions with Thierry Fournier and Jeevak Parpia.
We acknowledge the support from MICROKELVIN, the EU FRP7 low temperature infrastructure grant 228464.
\end{acknowledgments}


\end{document}